## ЭЛЕМЕНТАРНЫЕ ФИЗИКО-ХИМИЧЕСКИЕ ПРОЦЕССЫ

# NUMERICAL EVALUATION OF CUSPOID AND BESSOID OSCILLATING INTEGRALS FOR APPLICATIONS IN CHEMICAL PHYSICS

© 2004 г. J. N. L. Connor*, C. A. Hobbs**

*Department of Chemistry, University of Manchester, Manchester M13 9PL, UK

**Department of Mathematical Sciences, School of Technology, Oxford Brookes University, Wheatley Campus, Oxford OX33 1HX, UK

Oscillating integrals often arise in the theoretical description of phenomena in chemical physics, in particular in atomic and molecular collisions, and in spectroscopy. A computer code for the numerical evaluation of the oscillatory cuspoid canonical integrals and their first-order partial derivatives is described. The code uses a novel adaptive contour algorithm, which chooses a contour in the complex plane that avoids the violent oscillatory and exponential natures of the integrand and modifies its choice as necessary. Applications are made to the swallowtail canonical integral and to a bessoid integral.

## 1. INTRODUCTION

Oscillating integrals with coalescing saddle points often describe the scattering of atoms and molecules under short wavelength (or high frequency) conditions. They also arise in spectroscopic problems. Typical areas of application include the following:

• Rotational rainbows in atom-molecule and molecule-surface scattering.

• Analysis of vibrational transitions in molecular collisions.

• Theory of chemical reactions.

• Electron detachment in the scattering of negative ions.

• Charge transfer collisions.

• Penning ionisation.

• Pressure broadening of spectral lines and the appearance of satellite bands in spectra.

• Evaluation of Franck-Condon factors and the description of radiative transitions.

• Analysis of quasi-molecular orbital $X$-ray spectra.

• Atoms in magnetic fields and Stark spectroscopy.

• Autler-Townes effect for pulsed lasers.

More generally, oscillating integrals frequently arise in the theory of water, geophysical, electromagnetic and acoustic waves, as well as in the scattering of heavy nuclear ions. Many references to relevant research can be found in references [1–3].The evaluation of these oscillating integrals is often done using uniform asymptotic techniques [1, 2, 4]. Then an important problem is the numerical computation of certain canonical integrals and their first order partial derivatives. The simplest examples involve the cuspoid canonical integrals, which arise in the uniform asymptotic theory of oscillating integrals with 2, 3, 4, …, $n$ coalescing saddle points [1–4].

For two coalescing saddles, the canonical integral is the regular Airy function (or fold canonical integral):

$$\mathrm{Ai}(x) = (2\pi)^{-1}\int_{-\infty}^{\infty}\exp\left[i\left(\frac{1}{3}u^3 + xu\right)\right]du. \quad (1)$$

For three coalescing saddles, the canonical integral is the Pearcey integral (or cusp canonical integral) [5, 6]:

$$P(x, y) = \int_{-\infty}^{\infty}\exp[i(u^4 + xu^2 + yu)]du. \quad (2)$$

In the case of four coalescing saddles, the canonical integral is

$$S(x, y, z) = \int_{-\infty}^{\infty}\exp[i(u^5 + xu^3 + yu^2 + zu)]du, \quad (3)$$

which is called the swallowtail canonical integral [1–4].

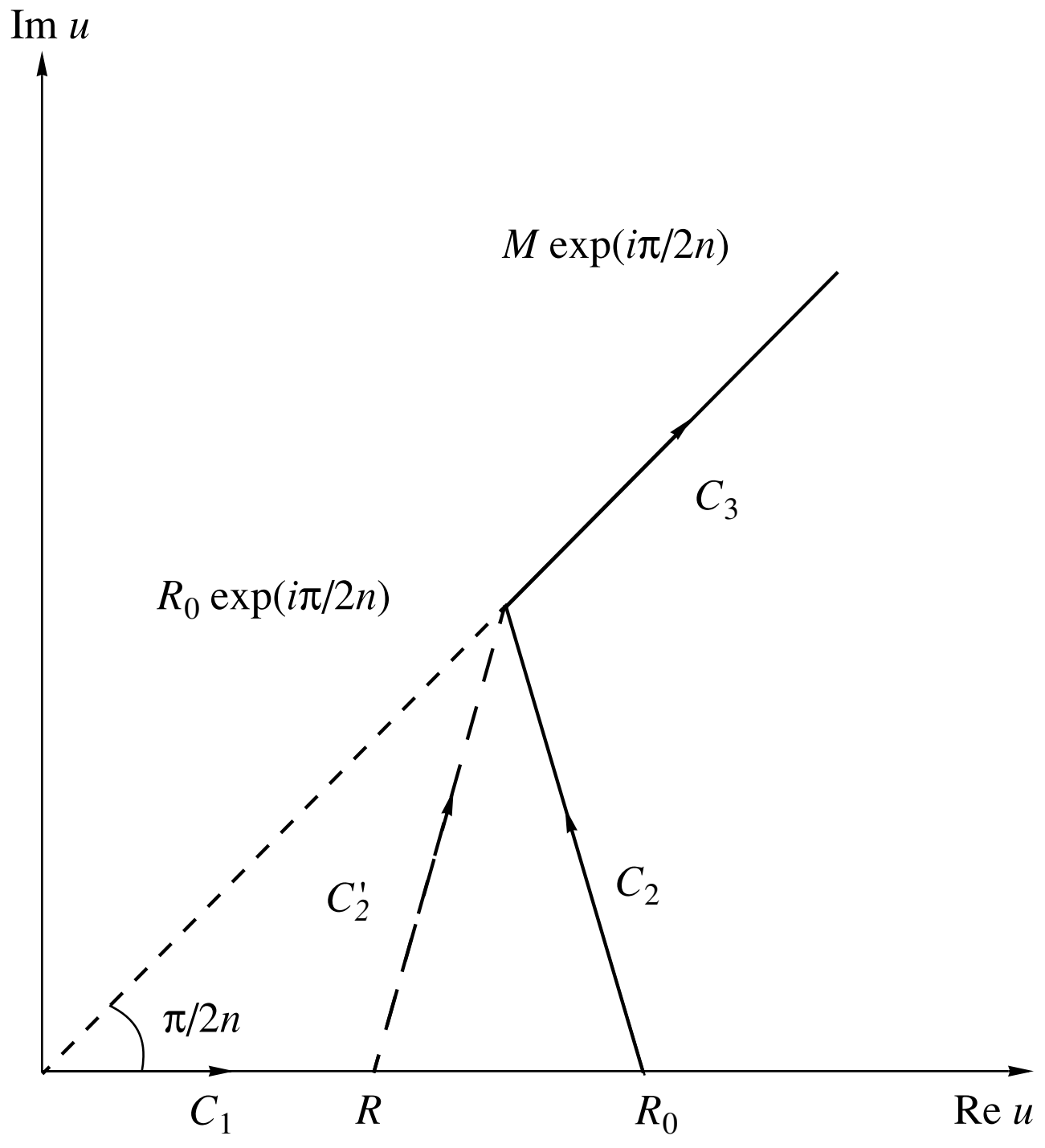

**Fig. 1.** Contours used for the numerical evaluation of $I_n^+(\mathbf{a})$. For simplicity, the superscript "+" has been omitted from $C_1^+$, $C_2^+$, $C'^+_2$, $C_3^+$, $R^+$, $R_0^+$ and $M^+$.

In the general case of $n - 1$ coalescing saddles, it is necessary to compute numerically the cuspoid canonical integral

$$C_n(\mathbf{a}) = \int_{-\infty}^{\infty} \exp[if_n(\mathbf{a}; u)]du, \qquad \mathbf{a} = (a_1, a_2, \ldots, a_{n-2}), \tag{4}$$

and its $n - 2$ first order partial derivatives

$$\frac{\partial C_n(\mathbf{a})}{\partial a_1}, \frac{\partial C_n(\mathbf{a})}{\partial a_2}, \ldots, \frac{\partial C_n(\mathbf{a})}{\partial a_{n-2}}$$

where

$$f_n(\mathbf{a}; u) = u^n + \sum_{k=1}^{n-2} a_k u^k,$$

with $u$ and the $a_k$ real and $n$ an integer greater than 2. The name of $C_n(\mathbf{a})$ arises because $f_n(\mathbf{a}; u)$ is a miniversal unfolding of a cuspoid singularity (also called an $A_{n-1}$ singularity) [1–4]. The importance of the integrals (1)–(4) can be seen from the fact that the forthcoming *Digital Library of Mathematical Functions* will contain a chapter entitled "Integrals with Coalescing Saddles" [7].

The purpose of this paper is to describe a computer code for the numerical evaluation of integrals of the type (1)–(4), and to present some illustrative results. The computer code is called CUSPINT and it uses an algorithm in which the integration path along the real axis is replaced by a more convenient contour in the complex $u$ plane, rendering the oscillatory integral more amenable to numerical quadrature.

Only an outline of the algorithm is presented since full details can be found in reference [3]. Numerical results are presented for $|S(x, y, z)|$. In addition, some results are shown for the bessoid integral [8]

$$J(x, y) = \int_0^{\infty} J_0(yu)u\exp[i(u^4 + xu^2)]du \tag{5}$$

where $x$ and $y$ are real and $J_0(\ldots)$ is the Bessel function of order zero. The bessoid integral (5) characterizes cuspoid focusing when axial symmetry is present [8].

Section 2 outlines three general methods for the numerical evaluation of cuspoid and bessoid integrals. The adaptive contour algorithm used in CUSPINT is described in section 3. Results for $|S(x, y, z)|$ and $J(x, y)$ are presented in section 4.

## 2. METHODS FOR THE NUMERICAL EVALUATION OF OSCILLATING INTEGRALS

There are three general methods available for the numerical evaluation of infinite oscillating integrals of the type (1)–(5).

***a) Maclaurin series.*** Maclaurin series expansions of (1)–(5) converge for all values of $x$, $y$, ... However, they are cumbersome to use when $|x|$, $|y|$, ... are large because of cancellation and slow convergence.

***b) Differential equations.*** This method derives a set of differential equations to which $C_n(\mathbf{a})$ or $J(x, y)$ is a solution and then solves the equations numerically. There are several advantages associated with this method: the derivatives are obtained automatically and it is an efficient way of generating grids of values for use in plotting. Disadvantages include: the derivation of the differential equations and their initial conditions is non-trivial; the method is difficult to implement on a computer for the case of general $n$; for certain values of $x$, $y$, ... the independent solutions of the differential equations are exponentially increasing, thereby limiting the accuracy to which the integrals can be calculated.

For $P(x, y)$ it is only necessary, in practice, to solve numerically one differential equation [9, 10] and this method has been used by Kaminski and Paris [11] to study the zeroes of $P(x, y)$ over a wide range of values of $x$ and $y$ (see also [12]). But for $C_n(\mathbf{a})$ with $n > 4$, the disadvantages of the differential equation method become serious.

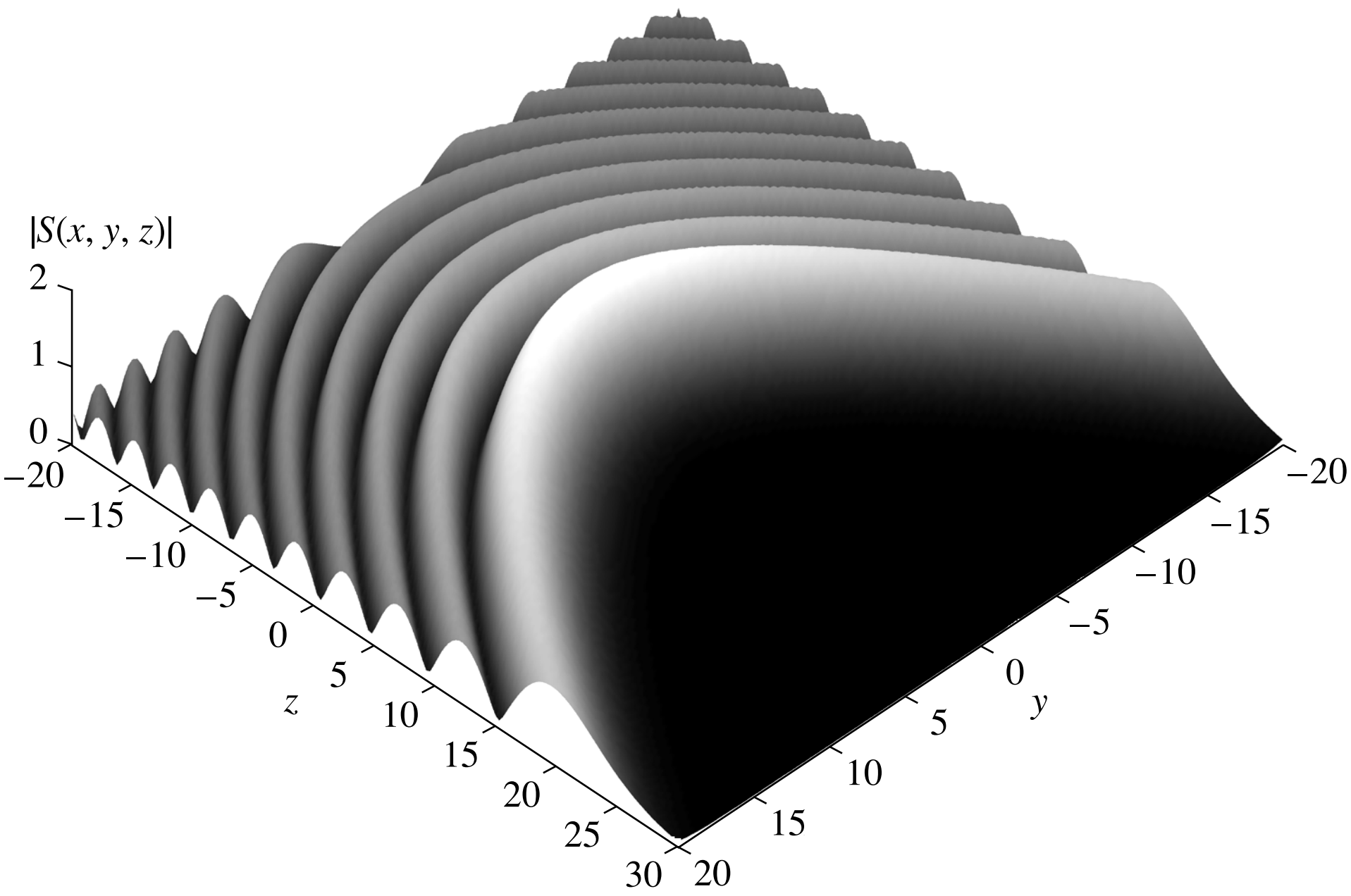

**Fig. 2.** Grey shaded perspective plot of $|S(x, y, z)|$ for $x = 4.0$.

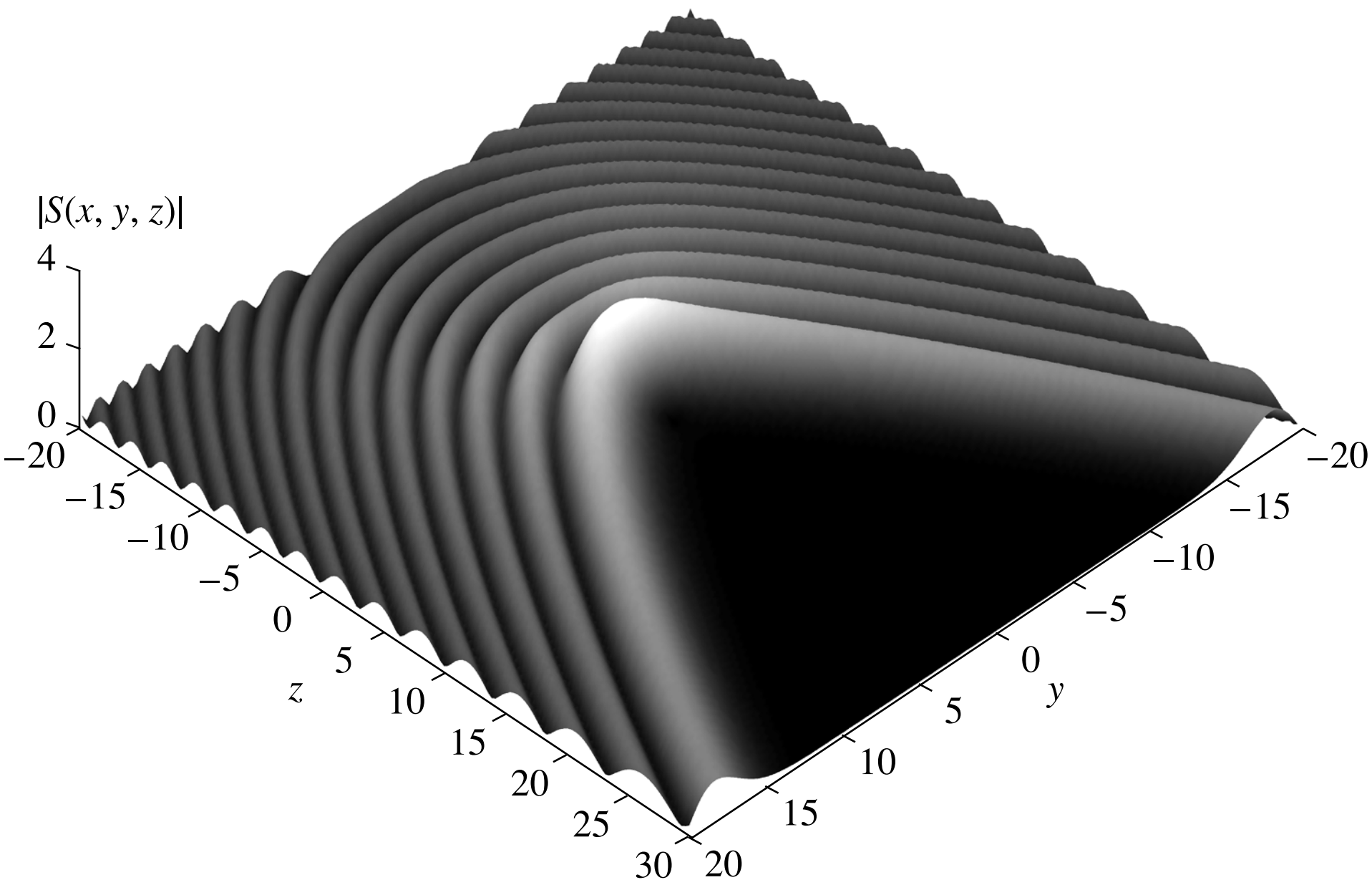

**Fig. 3.** Grey shaded perspective plot of $|S(x, y, z)|$ for $x = 0.0$.

***c) Contour Integral method.*** Since the integrand of $C_n(\mathbf{a})$ is infinitely oscillating along the real axis, a direct numerical evaluation is not possible. However, by deforming the contour of integration into the complex $u$ plane we can make the integrand more amenable to a numerical quadrature. This method has the advantage that it is efficient, gives high accuracy results, is relatively easy to implement on a computer and can be generalized to other types of oscillating integrals such as $J(x, y)$.

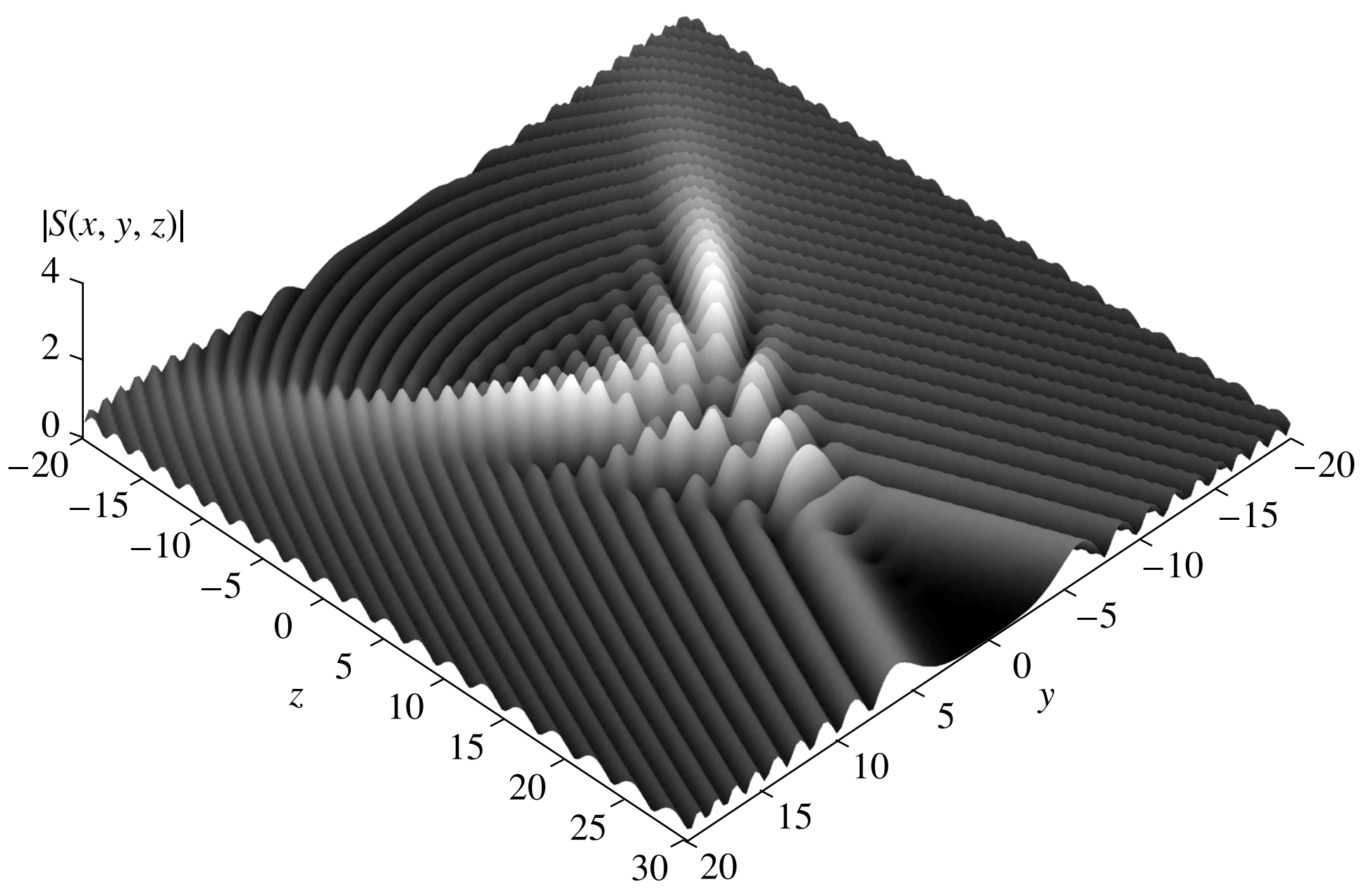

**Fig. 4.** Grey shaded perspective plot of $|S(x, y, z)|$ for $x = -6.0$.

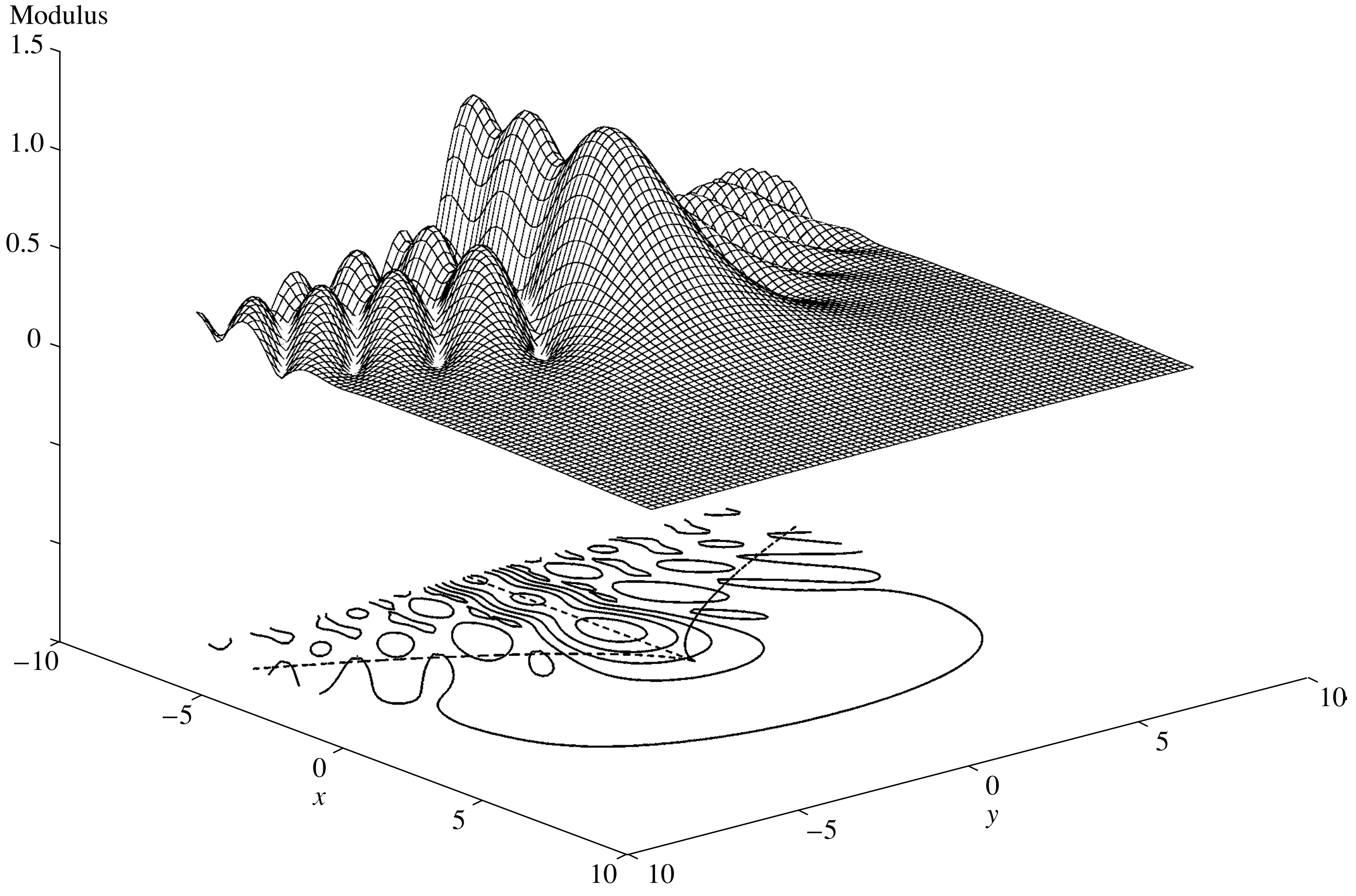

**Fig. 5.** Perspective and contour plots of $|J(x, y)|$. The broken curves are the branches of the caustic.

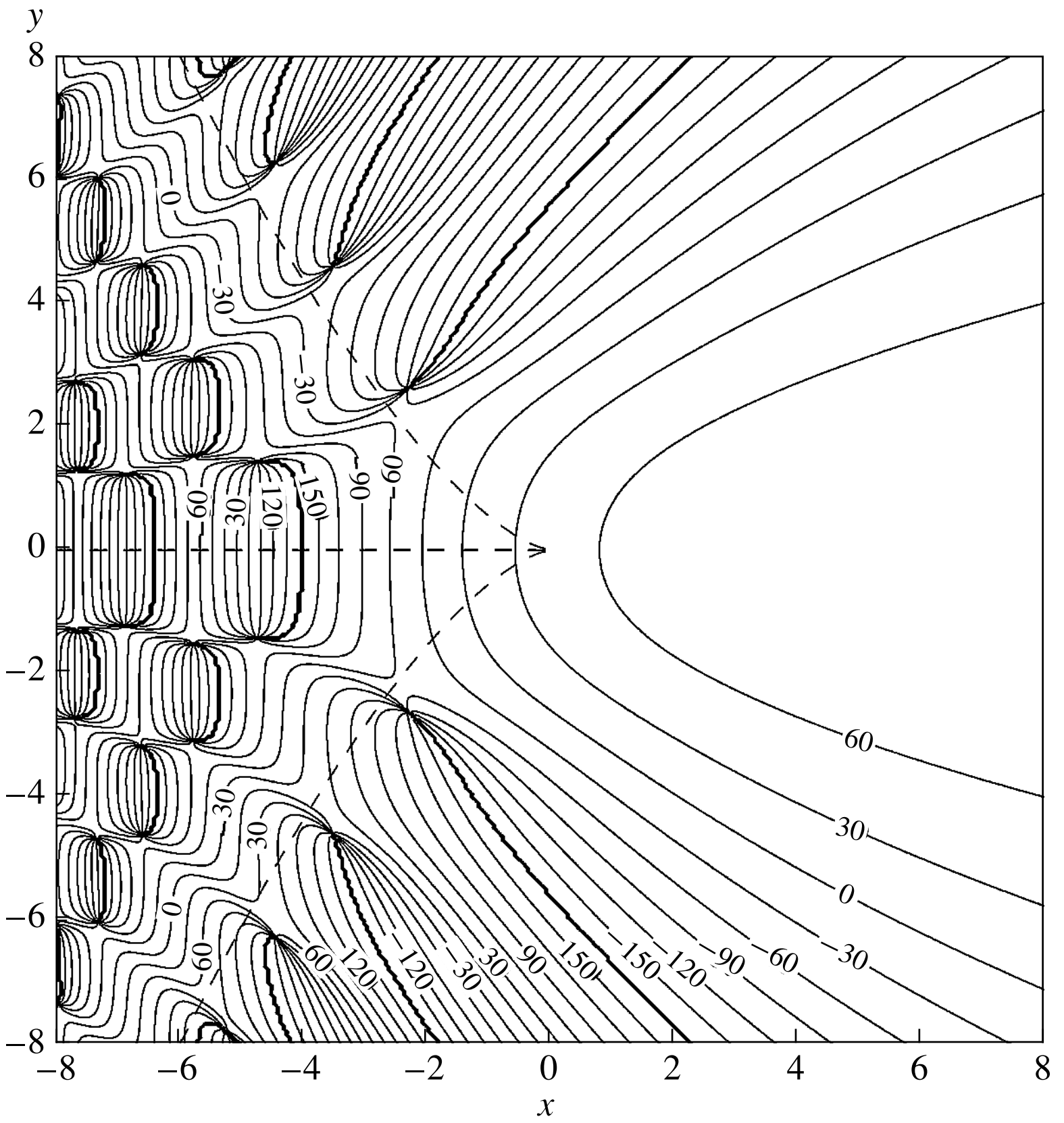

**Fig. 6.** Contour plot of arg$J(x, y)$/deg. The contours are –180(30)180. The thick full curves mark the phase discontinuities where arg$J(x, y)$/deg jumps in value from –180 to +180. The broken curves are the branches of the caustic.

## 3. ADAPTIVE CONTOUR METHOD

The first step is to write the general cuspoid integral (4) in the form

$$C_n(\mathbf{a}) = I_n^+(\mathbf{a}) + I_n^-(\mathbf{a}), \quad n = 3, 4, 5, \ldots$$

where

$$I_n^{\pm}(\mathbf{a}) = \int_0^{\infty} \exp[if_n(\mathbf{a}; \pm u)]du, \qquad \mathbf{a} = (a_1, a_2, \ldots, a_{n-2}). \tag{6}$$

We illustrate the adaptive contour method for $I_n^+(\mathbf{a})$, as the procedure for $I_n^-(\mathbf{a})$ is similar [3].

Next we observe that the integrand of equation (6) is infinitely oscillating along the real axis making a direct numerical evaluation impossible. However, we can use the ray from 0 to $\infty\exp(i\pi/2n)$ as a new contour of integration. This does not change the value of $I_n^+(\mathbf{a})$, as follows from an application of Cauchy's Theorem and Jordan's Lemma, using the fact that the integrand is entire. The new contour has the advantage that the integrand eventually becomes exponentially small, like $\exp(-t^n)$ with $t$ real, which suggests that a numerical approximation to $I_n^+(\mathbf{a})$ should be possible.

However, there is still a problem [13]: for certain values of the coefficients, $a_k$, the integrand can possess violent oscillations along the new contour, before it becomes exponentially small. This is a serious difficulty, which can prevent the accurate numerical evaluation of $I_n^+(\mathbf{a})$ along the direct ray from 0 to $\infty\exp(i\pi/2n)$ [13].

We solve the difficulties discussed above by a compromise in the choice of contours [3]. Figure 1 shows the three contours $C_1^+$, $C_2^+$, and $C_3^+$ that we employ.

The contour $C_1^+$ proceeds from the origin along the real axis to a *breakpoint*, $R_0^+$. The second contour $C_2^+$ is (usually) a straight line, which joins the point $R_0^+$ to the point $R_0^+\exp(i\pi/2n)$. The third contour $C_3^+$ lies along the original direct ray and joins $R_0^+\exp(i\pi/2n)$ to $M^+\exp(i\pi/2n)$ with $M^+ \geq R_0^+$. For suitable choices of $R_0^+$ and $M^+$, the infinite integral $I_n^+(\mathbf{a})$ can evidently be accurately evaluated, provided we can numerically compute the three finite integrals along $C_1^+$, $C_2^+$, and $C_3^+$.

The quadratures along $C_1^+$, $C_2^+$, and $C_3^+$ are performed in CUSPINT using specialist quadrature routines, especially suited to oscillating non-singular integrands. In particular, two versions of the code have been written: the first version uses the subroutine D01AKF present in the NAG Program Library [14], while the second version uses the subroutine DQAG and dependencies from the QUADPACK Program Library [15].

Reference [3] describes how CUSPINT chooses values for $R_0^+$ and $M^+$ for a given error tolerance. One problem that can arise is that the integral along $C_1^+$ is still oscillatory and difficult to evaluate numerically. To mitigate this problem, we can "cut the corner" and break away from the real axis at a point $R^+$, which is closer to the origin than is $R_0^+$. The integration along $C_2^+$ is then replaced by one along a new straight-line contour, $C'^+_2$, which joins $R^+$ to $R_0^+\exp(i\pi/2n)$ – see Fig. 1.

CUSPINT uses an iterative process which modifies the value of $R^+$ depending on the success, or otherwise, of the quadratures along $C_1^+$ and $C'^+_2$. This is possible because D01AKF and DQAG have powerful error detection facilities. CUSPINT acts on errors returned by the quadrature subroutines and modifies the contour of integration accordingly, *i.e.* an *adaptive contour algorithm* is used.

CUSPINT is written as a FORTRAN 90 module. Reference [3] describes in detail the input, output and error control flags, along with driver programs for test runs and the results that are produced.

## 4. APPLICATION TO THE SWALLOWTAIL AND BESSOID OSCILLATING INTEGRALS

This section presents results for two oscillating integrals: the swallowtail integral $S(x, y, z)$ and the bessoid integral $J(x, y)$.

Figures 2–4 show grey shaded perspective views of $|S(x, y, z)|$ for $x = 4.0$, 0.0 and –6.0 respectively. Each plot required 22378 numerical evaluations of $S(x, y, z)$ using the grid –20.0(0.3)29.8 for $z$ and –20.0(0.3)19.9 for $y$. The plots are symmetric about the line $y = 0$ because of the relation $S(x, -y, z) = S^*(x, y, z)$. The swallow's tail is clearly visible in Fig. 4. A detailed discussion of the structure in these plots can be found in references [10, 16] along with a description of the caustic surface, which acts as a "skeleton" upon which is built the "wave flesh" associated with $S(x, y, z)$.

Figure 5 shows a perspective and contour plot of $|J(x, y)|$. This plot required 6561 numerical evaluations of $J(x, y)$ on the grid –8.0(0.2)8.0 for $x$ and –8.0(0.2)8.0 for $y$. The corresponding contour plot for $\arg J(x, y)$/deg is displayed in Fig. 6. Both plots are symmetric about the line $y = 0$ because of the relation $J(x, y) = J(x, -y)$. Reference [8] contains a detailed discussion of the structure in the plots displayed in Figs. 5 and 6.

## ACKNOWLEDGMENTS

Support of this research by the Engineering and Physical Sciences Research Council (UK) and by INTAS (EU) is gratefully acknowledged.